# Moiré induced organization of size-selected Pt clusters soft landed on epitaxial graphene.

*Sébastien Linas[†,⊥], Fabien Jean[‡], Tao Zhou[§], Clément Albin[†], Gilles Renaud[§], Laurent Bardotti[†] and Florent Tournus[†,\*].*

[†]Institut Lumière Matière (ILM), Université de Lyon, UMR5306 Université Lyon 1-CNRS, 69622 Villeurbanne, France.

[‡]Institut Néel, CNRS and Université Joseph Fourier, BP166, F-38042 Grenoble Cedex 9, France

[§]Univ. Grenoble Alpes, INAC-SP2M, F-38000 Grenoble, France
CEA, INAC-SP2M, F-38000 Grenoble, France.

**Corresponding Author**

[\*]Email: <florent.tournus@univ-lyon1.fr>

**Present Addresses**

[⊥] Laboratoire des Multimatériaux et Interfaces (LMI), UMR 5615, Université Lyon 1-CNRS, 43 bd du 11 Novembre 1918, 69622 Villeurbanne, France

**ABSTRACT**

Two-dimensional hexagonal arrays of Pt nanoparticles (1.45 nm diameter) have been obtained by deposition of preformed and size selected $Pt_{80}$ nanoparticles on graphene. This original self-organization is induced, at room temperature, by the 2D periodic undulation (the moiré pattern) of graphene epitaxially grown on the Ir(111) surface. By means of complementary techniques (scanning tunneling microscopy, grazing incidence X ray scattering), the Pt clusters shapes and organization are characterized and the structural evolution during annealing is investigated. The soft-landed clusters remain quasi-spherical and a large proportion appears to be pinned on specific moiré sites. The quantitative determination of the proportion of organized clusters reveals that the obtained hexagonal array of the almost spherical nanoparticles is stable up to 650 K, which is an indication of a strong cluster-surface interaction.

**TEXT**

Supported metallic clusters have attracted a lot of interest in the last decades due to the dependence of their physical and chemical properties on their sizes and local environments.[1–6] Thus, a regular array of mono-dispersed clusters, providing a similar environment to each nanoparticle (NP), can highly facilitate their study or their use for optical,[7] magnetic[8,9] or catalytic[10–13] applications. Nanoparticles of typical lateral size and height of 1 to 2 nm are particularly relevant for the catalysis as they have been found to display much enhanced catalytic activity for this specific size.[12,14] One way to produce such arrays is the deposition or growth of NPs on templates in the form of patterned substrates. Among all possible templates, extensive studies have been performed on graphene epitaxially grown on several transition metals (g/metal).[15,16] As a matter of fact, moirés are often present. They arise from the lattice mismatch between the graphene and the metal surface,[17] and correspond to a regular 2D spatial variation of the corrugation and the electron density in graphene. Such a moiré super-lattice extends over the whole surface with a lattice unit cell parameter of a few nanometers (typically 2-3 nm). Additionally the properties of the graphene layer can be tuned by selecting the metal surface. For instance, in the case of Ru(0001),[18,19] graphene is strongly chemisorbed to the substrate whereas graphene is quasi freestanding when grown on Ir(111).[20–23] As a consequence, by combining the exceptional properties of graphene with the possibility to produce perfect moiré super lattice on transition metals,[24–26] this system appears as an ideal template for the fabrication of two-dimensional NP arrays. It has therefore been recently used to organize, by atomic vapor deposition (AVD), Ru,[27] Pt[28–30] and Co[31] clusters on graphene in epitaxy on Ru(0001) (g/Ru(0001)) as well as Ir,[32,33] Pt,[33,34] W[33] and Re[33] clusters on graphene in epitaxy on Ir(111) (g/Ir(111)). Nevertheless, the perfection of the cluster lattice strongly depends on the metal supporting the graphene film. Of the above mentioned, g/Ir(111) induces cluster arrays, the exceptional order and narrow size distribution of which makes them particularly attracting.[34] In the specific case of platinum clusters on g/Ir(111), three growth regimes have been found. First, a nucleation regime (up to 0.1 monolayer (ML) coverage) is characterized by a steep increase of the cluster density and by the presence of predominantly one atomic layer high clusters. For higher coverage (up to 0.75 ML), the cluster density becomes almost independent from the coverage and cluster size increases while almost all clusters evolve toward a height of two monolayers. For even higher coverages, the sintering taking place between neighboring clusters induces a progressive loss of organization.[33] Finally, since Pt clusters have a rather flat shape; the largest average size of ordered clusters is limited to 65 atoms.[35] Due to these specific growth regimes, the main limitations of platinum clusters produced by atomic deposition on g/Ir(111), regarding the catalytic applications for instance, are their small height of 1 to 2 MLs and the impossibility to tune independently cluster coverage and sizes.

The deposition of preformed spherical size-selected clusters appears as an original alternative technique to overcome these limitations.[36,37] A recent study, combining scanning tunneling microscopy (STM) observations and first principles calculations concluded to the possibility of moiré induced organization of extremely small $Pd_{n\ (n<20)}$ clusters on g/Ru(0001).[36] If the organization of small NPs is expected in a way similar to the organization of clusters formed by physical evaporation, the organization of much larger NPs on templates such as g/Ru(0001) or g/Ir(111) is expected to be more difficult, because of the much smaller ratio between interfacial NP atoms and bulk or surface ones. In the present work, we report on the investigation of a diluted two dimensions (2D) array of preformed and size selected $Pt_{80}$ clusters (1.45 nm mean diameter) that are soft-landed at room temperature (RT) on g/Ir(111) by means of three complementary techniques: STM, grazing incidence X-ray diffraction (GIXD) and grazing incidence small-angle X-ray scattering (GISAXS). The morphological and structural evolutions of this system are also investigated during annealings above RT. An efficient and original method to quantify the organization order of the cluster super-lattice at a macroscopic scale is proposed.

**RESULTS AND DISCUSSION**

Typical STM images of size-selected Pt clusters deposited on g/Ir(111) are shown in Figure 1a,b for two incident NP densities. They reveal isolated particles the density of which corresponds to the incident one, evidencing the absence of coalescence and confirming the absence of fragmentation of the supported clusters. Moreover, neither cluster aggregation nor specific step decoration is detected, indicating that clusters do not diffuse over distances larger than a few nanometers. Note that, this low diffusion is in complete contrast with the case of Pt clusters deposited on graphite, where NPs diffuse rapidly and gather to form "island" of NPs.[38] This limited diffusion of platinum NPs on g/Ir(111) suggests their pinning on the surface. Such a pinning could result from adsorbed species on the g/Ir(111) surface (e.g. carbonaceous material[37,39]), defects of the graphene lattice [20,40] or anchoring of the NPs on specific sites of the moiré pattern. Nevertheless, immobilization of NPs by adsorbed species has not been observed in the case of Pt clusters deposited on graphite.[41] Such a contamination is unlikely here since all samples were kept under ultrahigh vacuum (UHV) and is additionally confirmed by STM atomic resolution observations.[42] Concerning the trapping of clusters by natural defects, according to the extremely high quality of graphene performed on Ir(111),[24] a pure heterogeneous growth on defects appears unlikely. On the contrary, according to the moiré specific site density ($1.85 \times 10^5$ sites.$\mu m^{-2}$, assuming one pinning site per moiré unit cell), the observed supported cluster density rather favor their pinning on the moiré pattern. The NP height histogram measured by STM is shown in Figure 1c. The average NPs height (<H> = 1.45 ± 0.06 nm) is close to the average diameter (<D>) of the incident clusters, suggesting that the supported NPs conserve their spherical shape when going from the free (*i.e.* clusters beam) to the supported state. This last result is drastically different from those obtained previously by AVD where, as mentioned before, supported clusters are rather flat (typically 1 to 3 monolayers high).[35] For the low surface coverage the moiré of g/Ir(111) as well as the clusters are clearly visible in STM images as in the AVD studies. Unfortunately the convolution between the tip and the clusters, enhanced by the 3D morphology of the supported NPs, prevents the precise determination of their structural relationship with the surface and their location on the moiré lattice. To get more information on this last point, the nearest neighbor distance $d_{NN}$ (center to center) of Pt clusters is shown in Figure 1d for a medium density. The $d_{NN}$ histogram can be fitted by two Gaussian functions centered on $d_1 = 2.4$ nm and $d_2 = 2.4 \times \sqrt{3}$ nm. This shows that, despite the randomness of the cluster deposition process,[43] the final locations of Pt clusters on g/Ir(111) are not randomly distributed. Additionally the mean $d_{NN}$ distance is clearly compatible with the pinning of the Pt NPs on specific sites of the g/Ir(111) moiré. However, we emphasize at this point that this analysis cannot exclude the presence of some randomly distributed clusters. Such a feature could explain the absence of a clear cluster organization on the STM images for this surface coverage.

Furthermore, STM does not provide precise information on several characteristics of this system, e.g. the epitaxial relationship of the NPs with the substrate, their precise shape or the correlation length of the NPs organization. GIXD and GISAXS studies were then conducted to answer these questions.

In-plane scans of the scattered X-ray intensity were recorded along the *h* direction in the vicinity of the (200) rod of Ir (Figure 2a) for g/Ir(111) before and after NP deposition. The (200) rod of Ir is centered at *h* = 2, the (200) Bragg peak of graphene at $h \approx 2.23$ and two moiré peaks can be observed around $h \approx 1.9$ and $h \approx 2.1$.[26] The unit cell parameter of the moiré can be deduced from the relative positions of the (110) Bragg peaks of Ir and graphene (not shown), according to: $a_{\text{moiré}} = (a_{Ir}^{-1} - a_{Graphene}^{-1})^{-1} = 2.48$ nm. This result is in complete agreement with previous studies[26,32] and with the present STM observations. The deposition of NPs induces two main effects at RT on the in-plane scans (Figure 2a), (i) a broadening of the Ir rod and (ii) the raise of the moiré peak around $h \approx 1.9$ (see inset of Figure 2a).

Because the nearest neighbor distances of Pt (2.77 Å) is very close to that of Ir (2.72 Å), NPs in epitaxy on the Ir(111) surface are expected to yield diffraction peaks in the vicinity of the Ir peaks and rods. These peaks of epitaxial NPs are expected to be broad if they are randomly distributed on the surface, and to be peaked at the moiré, Ir and graphene rod positions if they are organized on the moiré. If the NPs are not in epitaxy but adopt many orientations, the scattered intensity is distributed over a portion of sphere and thus the intensity measured in a single radial scan such as that of Figure 2a should be negligible. Measurable intensity from the NPs along such scans is expected only if a significant fraction of the NPs are in epitaxy with the substrate. Consequently, the much wider peak appearing around the (200) rod upon NP deposition shows that some NPs are in epitaxy with Ir(111). Another observation is the increase of the moiré peak intensity, which implies that some NPs are pinned on a specific site of the moiré cell. Hence, a qualitative analysis by GIXD indicates that a significant fraction of the NPs are in epitaxy with the Ir(111) substrate; and at least part of them are perfectly anchored to a specific moiré site; the rest possibly being in different places (*i.e.* with no coherence from one unit cell to the other). We then define Θ as the proportion of clusters pinned on the moiré.

The indications of the organization and epitaxy of the clusters given by GIXD have been followed during an annealing of the samples. In Figures 2 b and c, the vicinity of the (200) rod of Ir and the moiré peak around $h \approx 1.9$ are reported for the same sample at several annealing temperatures. After annealing at 460 K, the width of the (200) rod of Ir is reduced to the value found for the bare g/Ir(111). The intensity of the moiré peak around $h \approx 1.9$ remains almost constant until T = 550 K and then decreases and reaches the background level at T = 650 K.

At this point, two preliminary conclusions can be drawn. Between RT and 460 K the clusters that were in epitaxy but are not positioned on specific sites have lost their epitaxial relationship with Ir(111). The decrease of the moiré peak with temperature above 560 K (see Figure 2c) suggests either a loss of epitaxy or a decrease of Θ or a combination of these two features.

To obtain more quantitative information on the cluster shape and organization, the scattered intensity has been measured near the origin of the reciprocal space using the GISAXS technique.[44,45]

The scattering cross section at small angles [$I(\mathbf{q})$, where **q** is the scattering wave vector] can be expressed, in terms of NPs' form factors ($F^i \equiv \int_{S^i} e^{-i\mathbf{q}\cdot\mathbf{r}} d^3r$, where $S^i$ is the shape function) and the position of each NP on the surface ($\mathbf{R}_\parallel^i$) according to:

$$I(\mathbf{q}) = \sum_i |F^i(\mathbf{q})|^2 + \sum_{i,j\neq i} F^i(\mathbf{q}) F^{j,*}(\mathbf{q}) e^{-i\mathbf{q}\cdot(\mathbf{R}_\parallel^i - \mathbf{R}_\parallel^j)} \quad (1)$$

The scattered intensity at small angles contains therefore information about the shape of the NPs and the spatial organization of the particles. In a GISAXS pattern, a correlation peak appears in addition to the form factor if the clusters are spatially organized. In Figure 3a (left panel) a map of the reciprocal space for $2\theta < 100$ mrad and at $\alpha_f = \alpha_c = 7$ mrad is shown for g/Ir(111) covered by Pt NPs. This map is a result of collecting GISAXS patterns at different azimuths and for which the intensity at the Yoneda peak[46] ($\alpha_f = \alpha_c = 7$ mrad) has been extracted. One can see in figure 3b that a correlation peak is only visible in GISAXS patterns measured when the incident beam is aligned with the [100] (left panel) and [110] (center panel) Ir crystallographic orientations. As was shown for Co clusters organized on the 2D-patterned Au(111)[47] and Au(677)[48] surfaces, this in-plane reciprocal space map demonstrates the presence of a two-dimensional order of the Pt NPs. The corresponding unit cell of this 2D array is hexagonal with two sides aligned along the Ir[100] and the Ir[010] directions and a lattice parameter of $d_{Pt\,array} = 2.485\,nm$. The sharpness and the dependence of the correlation peak with crystallographic orientation show that the organization of the clusters is not only due to a favored first neighbor distances as is in the case of Pt NPs deposited on graphite,[41] but clearly reflect the pinning of the clusters on specific sites. We can conclude from these results that some of the Pt NPs are pinned on the hexagonal moiré of g/Ir(111).

Additional characteristics of the 2D array of Pt clusters can be extracted by fitting the GISAXS patterns, using the IsGISAXS[45] software. $I(q)$ is assumed to be the incoherent sum of the intensity scattered by NPs lying at random locations ($I_R$) and that scattered by the NPs anchored on the graphene moiré ($I_L$):

$$I(q) = \Theta I_L(q) + (1 - \Theta)I_R(q) \qquad (2)$$

The usual[46] approximation to simulate the GISAXS intensity of an assembly of nanoparticles is the local mono-disperse approximation (LMA) which assumes that all NPs locally have the same size; the size variation arising only over long distances. However, in the present case, we have a random deposition of NPs having a pre-defined size distribution and the 2D super-lattice is randomly occupied by preformed cluster, which may slightly vary in size from site to site. The experimental system studied in this work is thus a model system for the decoupling approximation[49] (DA) in which the size of each NP is independent of its neighborhood. In this framework, the scattered intensity is the sum of two terms:[45,46]

$$I_m(q) = \{\langle|\mathcal{F}(q)|^2\rangle_D - |\langle\mathcal{F}(q)\rangle_D|^2\} + |\langle\mathcal{F}(q)\rangle_D|^2 S_m(q); \quad m = L, R \qquad (3)$$

where $\mathcal{F}$ is an effective form factor[45] calculated within the distorted-wave Born approximation and $\langle...\rangle_D$ denotes averaging over the size distribution. $S(q)$ is the total interference function which describes the statistical distribution of the Pt clusters on the surface and thus their lateral correlations:

$$S(q) = 1 + \rho_S \int d^2 R_{ij} g(\mathbf{R}_\parallel^{ij}) e^{-i\mathbf{q}\cdot\mathbf{R}_\parallel^{ij}} \qquad (4)$$

where $g(r)$ is the lateral 2D particle-particle pair correlation function. In the case of this experiment, we then have [$S(q) = 1$ for a random distribution]:

$$I_L(\boldsymbol{q}) = \{\langle|\mathcal{F}(\boldsymbol{q})|^2\rangle_D - |\langle\mathcal{F}(\boldsymbol{q})\rangle_D|^2\} + |\langle\mathcal{F}(\boldsymbol{q})\rangle_D|^2 S_L(\boldsymbol{q}) \qquad (5)$$

$$I_R(\boldsymbol{q}) = \langle|\mathcal{F}(\boldsymbol{q})|^2\rangle_D \qquad (6)$$

In order to extract quantitative parameters from GISAXS measurements, parallel and perpendicular line cuts, obtained from a GISAXS pattern (see Figure 3c,d) are fitted using the following procedure. First, the cluster morphological parameters (<D>, <H>, $\sigma_D$) are determined from an out of azimuth direction assuming that the NPs are truncated spheres. Then, this set of parameters is used to fit the correlation peak around the Ir[100] direction. This gives access to the characteristic length (C.L.) of the exponentially decaying correlation of the 2D Pt cluster array. Note that $S_L(\boldsymbol{q})$ is considered to correspond to a perfect 2D hexagonal lattice (reflecting the moiré pattern), with a lattice parameter adjusted to reproduce the peak positions in the GISAXS map (*i.e.* $d_{Pt\,array} = 2.485\,nm$). Finally, $\Theta$ can be determined from the correlation peak intensity, with respect to the background. For this purpose, a line cut from an off-azimuth GISAXS pattern (*i.e.* where $S_L(\boldsymbol{q})$ is simply equal to 1) is subtracted to the corresponding line cut along the Ir[100] direction. The resulting curve is then proportional to the $S_L(\boldsymbol{q})$ difference and only the correlation peak remains. The proportion of organized Pt clusters is thus deduced from a set of simulations at different $\Theta$ values (Figure 4a). In the end, as shown in Figure 3, an excellent agreement between the experimental data and the simulations is found, not only for representative line cuts but also for the entire GISAXS patterns and azimuthal map (Figure 3).

The morphological parameters found for the best fit of the GISAXS data are <D> = 1.51 nm, <H> = 1.25 nm and a Gaussian dispersion of standard deviation $\frac{\sigma_D}{\langle D \rangle} = 11.6\,\%$. The shape of the NPs derived from GISAXS is almost spherical ($\frac{D}{H} \approx 0.83$) with a small wetting of the substrate. Note this shape value is not exactly the same as that deduced by the STM, which might arise from the geometrical model of the NPs (truncated sphere) which dismisses the true atomic arrangement and in particular the faceting. The size of the graphene domains (*ca.* 62 nm), calculated from the GIXD graphene peaks,[26] is close to the correlation length (C.L. $\approx 92$ nm), deduced from the GISAXS correlation peak. This suggests that the cluster hexagonal organization is limited by the size of the graphene domains. The proportion of clusters pinned on the moiré is found to be $\Theta = 50\pm2\%$ for a medium density of clusters. Note that, for the sample covered with a high density of clusters (see Figure S1 in the supporting information) the results are very similar ($\Theta = 50\%$). We deduce that half of the preformed Pt clusters deposited on g/Ir(111) self-organize as an incomplete (*i.e.* with vacancies) 2D lattice of clusters, the rest being off lattice. Let us emphasize that the occupation ratio of the moiré sites are 8 and 27% for the samples covered with a medium and a high densities of NPs.

As the temperature increases (RT to 650 K) the intensity of the correlation peak clearly decreases (Figure 4b): $\Theta$ is found to decrease (Figure 4c) from 50 to 42% without change in cluster shape and size. It can be noticed that the 2D array is relatively stable upon high temperature annealing since at 650 K only 16% of the clusters previously pinned by the moiré have left the super-lattice. Between T = 556 and 560 K the value of $\Theta$ determined by GISAXS barely evolves while the moiré peak in GIXD (Figure 2c) drastically decreases: this can be interpreted as a loss of epitaxy with the Ir(111) surface, although the clusters are still anchored at specific moiré site. This emphasizes the strength of the GISAXS technique which is sensitive to all NPs whether they are in epitaxy or not.

For temperatures higher than 650 K, the shape of the clusters is modified (see Figure S2 in supporting information). These morphological modifications are probably due to the coalescence or sintering of NPs, or possibly to intercalation of Pt below the graphene. Anyway the system is no more a diluted 2D array of size selected clusters and the modification is irreversible as confirmed by measurements after cooling down to RT.

The stability of the cluster arrays with temperature indicates a strong bonding of the NPs to the substrate. However the decrease of $\Theta$ suggests that the clusters leaving the specific moiré sites are trapped elsewhere. Two explanations can be drawn to account for the out of lattice clusters: (i) in some areas of the sample the clusters are fully organized whereas on other areas clusters are not pinned on the moiré specific sites. This could be the results of a non-complete coverage of Ir(111) by graphene, or trapping of the clusters by wrinkles that form in g/Ir(111).[50] However, the homogeneity observed from STM images of various areas seems to refute this hypothesis. (ii) The cluster adsorption energy landscape could present one principal potential well, able to attract clusters in an area corresponding to *ca.* half the super-lattice unit cell, while there exists many smaller potentials wells distributed over the rest of the unit cell. A theoretical study should help understanding the kinetics and thermodynamics of NP trapping on g/Ir(111). This could eventually give clues to enhance the ratio of organized clusters, for instance by increasing the temperature during the deposition process.

**CONCLUSIONS**

In summary, we have shown that about half of the incident Pt clusters soft-landed at RT organize on a 2D array templated by the moiré of g/Ir(111). The striking agreement between experimental and simulated GISAXS patterns enables a first quantitative characterization, on a macroscopic length-scale, of the organization of quasi-spherical Pt nanoparticles. From GIXD measurements, we have found that some of the NPs are in epitaxy with the Ir(111) substrate. Additionally, a moderate annealing (at 460 K) suppresses the epitaxial relationship, while the particle shape and size remain unchanged. Besides, as deduced from the GISAXS correlation peak, the organization shows minor variations up to 650 K. This demonstrates that the obtained 2D arrays of NPs are stable against quite high temperature annealing: 84% of the clusters that were organized at RT are still anchored on specific moiré sites at 650 K. Our findings provide a first step in the investigation of self-organized 2D arrays of preformed clusters deposited on a graphene surface. The preparation of well-defined Pt nanoparticles samples can then be promising for catalytic applications. Furthermore, our original approach could also be extended to bi-metallic and/or magnetic materials which cannot be organized on graphene by atomic deposition. Indeed, the clusters obtained by AVD are the result of a nucleation and growth process occurring on the surface. On the moiré of

g/Ir(111), cluster super-lattices can then be formed for high cohesive energy metals, as Ir, Pt or W, whereas lower cohesive energy materials, such as Au, Fe or Ni,[33] do not organize. The mass-selected low energy cluster beam deposition (MS-LECBD) technique, where preformed clusters having a 3D morphology are soft-landed on a surface, could overcome this limitation.

**METHODS**

**Preparation of g/Ir(111).** The substrates preparation was performed in the UHV chamber (base vacuum $5\times10^{-11}$ mbar)[51] of the BM32 beamline of the European Synchrotron Radiation Facility[52] (Grenoble, France). Two different Ir single crystals, cut and polished on a (111) surface termination within 0.1° were used (one was provided by Surface Preparation Laboratory and the other by Mateck). They were cleaned by cycling high temperature annealing periods (1473 K) and ion bombardment (1.3 kV/10 µA Ar+ for 30 min at RT), with a final annealing in oxygen ($5\times10^{-7}$ mbar for 10 min at 1300 K). The graphene was prepared following a well-established method, consisting of a temperature programed growth step (adsorption of ethylene at RT followed by a flash annealing at 1473 K) and a chemical vapor deposition step (annealing at 1273 K under an ethylene partial pressure of $10^{-7}$ mbar).[24,25]

The g/Ir(111) substrates were then transferred in a homemade UHV transfer device ($10^{-8}$ to $10^{-9}$ mbar) from the BM32 beamline to the nearby cluster deposition chamber (PLYRA facility[52]). The transfer typically lasted a few hours.

**Clusters synthesis.** Pt clusters have been synthesized by the mass-selected MS-LECBD technique described elsewhere.[37,53] Briefly, the NPs have been first produced by a laser vaporization source and subsequently mass-selected using a quadrupolar electrostatic deviator.[53] In the present experiment, the deviator was set to produce incidents clusters with a diameter of *ca.* <D> = 1.45 ± 0.06 nm.[37] The clusters were then soft landed at RT on a g/Ir(111) under UHV ($10^{-11}$ to $10^{-10}$ mbar). The incident flux of NPs (cations) has been determined using a Faraday cup connected to a pico-amperemeter. With this technique, the density and size of the clusters are set independently. In this work low ($10^4$ NPs μm$^{-2}$), medium ($3\times10^4$ NPs μm$^{-2}$) and high ($10^5$ NPs μm$^{-2}$) densities of incidents clusters were deposited (relative uncertainties of 8 %). The deposition chamber is connected to an Omicron UHV STM allowing *in situ* STM observation of the samples at RT. After NPs deposition, the samples were transferred back under UHV to the synchrotron facility in Grenoble (France).

**GIXD and GISAXS measurements** with synchrotron light were performed in an UHV chamber ($5\times10^{-11}$ mbar) coupled with a Z-axis diffractometer at the ESRF BM32 beamline.[51,52] The monochromatic photon beam had an energy of 11 keV, a focus size of 300×300 (H×V) μm$^2$ and an incidence angle of $\alpha_i$ = 0.38° (just below the critical angle for total external reflection of Ir which is 0.42° at 11 keV). The intensity scattered by the surface was collected at wide angle (GIXD) using a 2D pixel detector (Maxipix). Slits (0.5 mm, parallel to the surface) have been placed well before the detector (slit-sample distance of 190 mm, compared to a detector-sample distance of 640 mm). The reference for the normalized scattered intensities (NSIs) is the incident beam whose intensity has been measured by a monitor placed before the sample. The intensity scattered at small angle (GISAXS) was collected using a high grade charge-coupled device 16-bit camera (Photonic Science; 2.25 Mpixels pixel size of 43.965 μm$^2$). The sample to detector distance was set to 525±5 mm. The *h*, *k*, *l* indexes in relative lattice unit are referred to the Ir(111) hexagonal surface unit cell ($a_S = b_S$ = 2.715 Å; $c_S$ = 6.65 Å).

**ACKNOWLEDGMENT**

We thank O. Boisron (ILM), O. Geaymond, and the staff of the BM32 beamline. Research supported by French ANR Contract No. ANR-2010-BLAN-1019-NMGEM.

**Author contribution**

L.B., F.T. and G.R. initiated the study. F.J. and T.Z. performed the graphene synthesis; S.L., C.A., F.T. and L.B. involved in the clusters synthesis and deposition; T.Z., F.J., G.R., S.L. and F.T. performed the x-rays measurements; L.B. performed the STM imaging; C.A. designed and build the UHV transfer device. S.L. wrote the manuscript and prepared the figures. T.Z., F.J., G.R., S.L., L.B. and F.T. participated to the scientific discussion, analysed the data, reviewed the manuscript and the figures.

**ADDITIONAL INFORMATION**

**Supporting Information available:** accompagnies this paper at http://www.nature.com/scientific reports

**Competing financial interests:** The authors declare no competing financial interests.

**Figures**

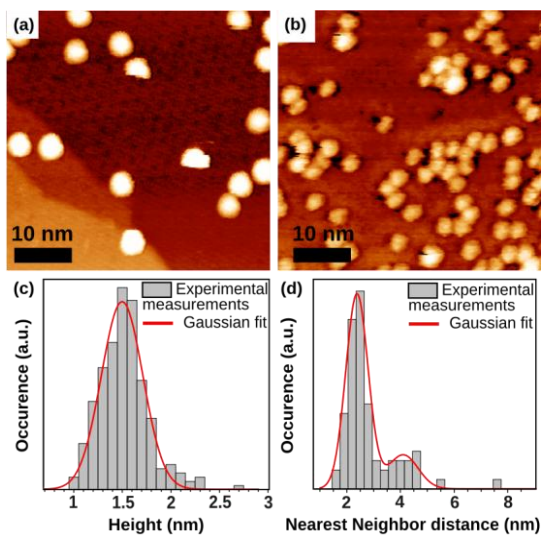

**Figure 1.** (a-b) Typical STM topographs of size-selected (1.5 nm in diameter) platinum clusters supported on g/Ir(111). The coverage of the NPs corresponds to a low (a) and medium (b) density. (c) Height distribution of the Pt clusters measured by STM, fitted by a Gaussian function. (d) Nearest neighbor distance (center to center) distribution of Pt clusters deposited on g/Ir(111) with a medium density. The distribution is fitted by two Gaussian functions.

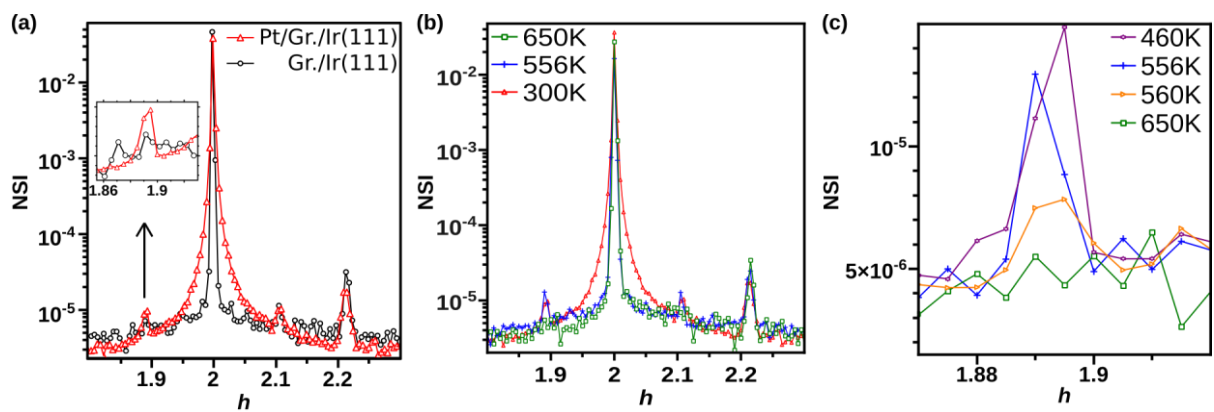

**Figure 2.** In-plane scans along the *h* direction in the neighborhood of the (200) Ir rod. The temperatures are (a) RT, (b) RT to 650 K and (c) 460 to 650 K. The samples were (a), bare g/Ir(111) (black circles) and g/Ir(111) covered with a medium density of Pt clusters (red triangle) and (b, c) g/Ir(111) covered with a medium density of Pt clusters. In (a), the inset shows a zoom of the scan in linear scale.

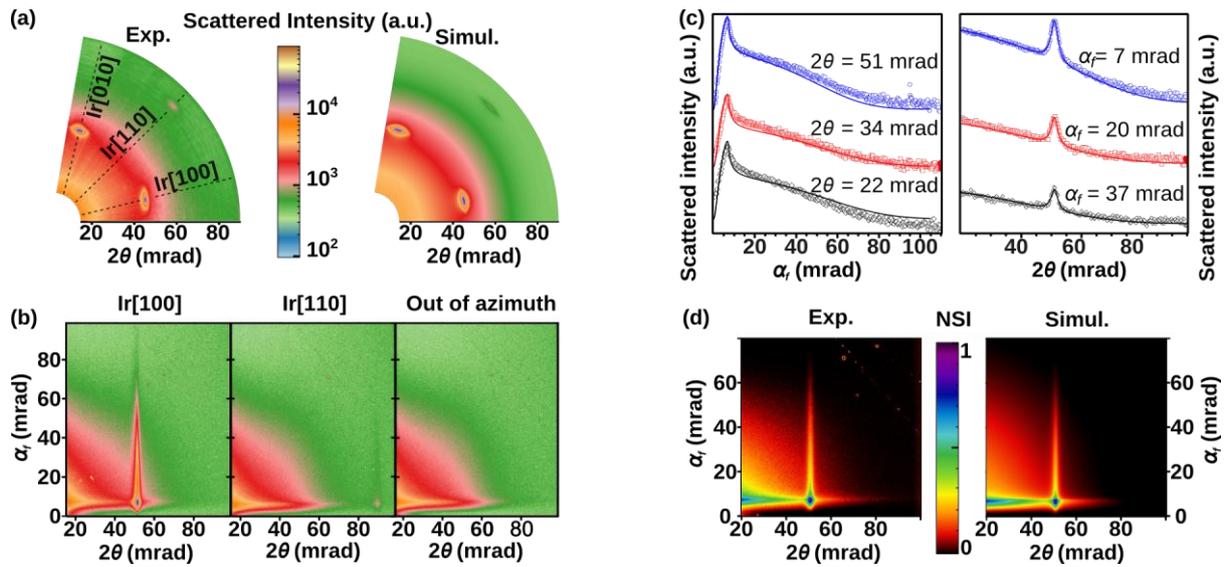

**Figure 3.** (a) Map of the reciprocal space at RT at $\alpha_f = \alpha_c = 7$ mrad (left) and the corresponding simulation (right). (b) GISAXS patterns at RT, with the incident beam along the Ir[100] (left), Ir[110] (center) directions and out of azimuth (right). (c) Perpendicular (left) and parallel (right) line cuts of the 2D GISAXS pattern shown in (d), the continuous lines correspond to the best fits and the symbols (circles, squares and diamonds) correspond to the experimental data points. The curves have been shifted for clarity. (d) 2D experimental GISAXS pattern at RT with the incident beam along the Ir[100] direction (left) and the corresponding simulation (right). (a) and (b) correspond to a sample of Pt clusters deposited on g/Ir(111) with a high density (in order to have a better visibility of the correlation peak in the direction Ir[110]), while (c) and (d) correspond to a medium cluster density.

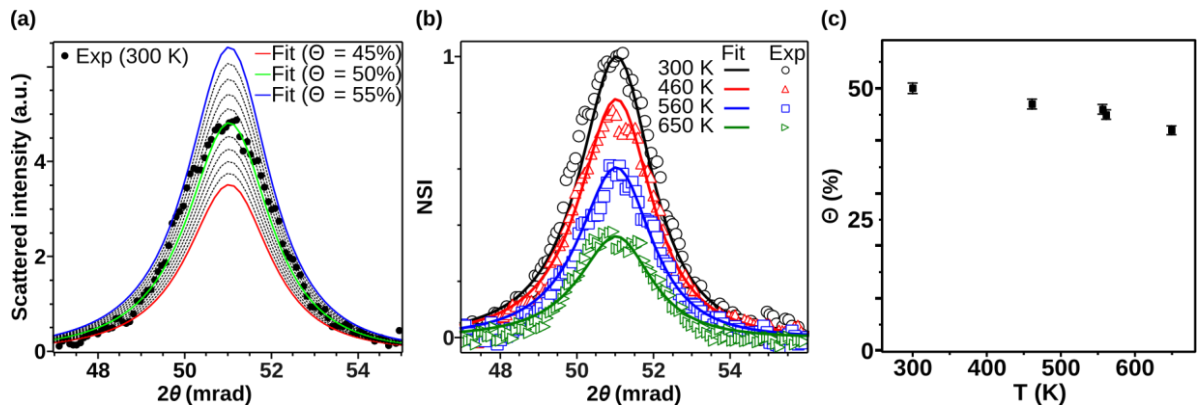

**Figure 4**. (a) Fit of the RT experimental GISAXS correlation peak (black dots) with various fraction Θ of Pt clusters lying on the moiré lattice. Difference of 2D GISAXS pattern line cuts ($α_f = α_c ≈ 7$ mrad) with the incident beam 10° off azimuth and in the Ir[100] direction. The intermediates values of Θ are represented in dashed black lines, with an increment of 1%. (b) Best fits of experimental GISAXS correlation peaks at various temperatures. (c) Plot of Θ versus the annealing temperature. Pt NPs have been deposited on g/Ir(111) with a medium density.